\documentclass[twocolumn,english,aps,pra,tightenlines,longbibliography]{revtex4-2}
\usepackage{lmodern}
\usepackage{amsmath}
\usepackage{amssymb}
\usepackage{graphicx}

\makeatletter
\usepackage{microtype}

\usepackage[unicode=true,pdfusetitle,
 bookmarks=true,bookmarksnumbered=false,bookmarksopen=false,
 citecolor=cyan,urlcolor=magenta,linkcolor=blue, breaklinks=false,
 pdfborder={0 0 0},backref=false,colorlinks=true]{hyperref}

\usepackage{amsthm}\usepackage{epstopdf}\usepackage{color}
\usepackage{bm}\usepackage{babel}
\usepackage{upgreek}
\usepackage{textcomp}
\usepackage{siunitx}
\newcommand\figref[2]{Fig.\,\ref{#1}\hyperref[#1]{({#2})}}
\newcommand\figsref[2]{Figs.\,\ref{#1}\hyperref[#1]{({#2})}}
\newcommand\fref[2]{\ref{#1}\hyperref[#1]{({#2})}}

\makeatother

\usepackage{babel}

\begin{document}
\title{Exploiting the combined dynamic and geometric phases for optical vortex
beam generation using metasurfaces }
\author{Jialong Cui, Chen Qing, Lishuang Feng, and Dengke Zhang}
\email{dkzhang@buaa.edu.cn}

\affiliation{School of Instrumentation and Optoelectronic Engineering, Beihang University, Beijing 100191, China}
\begin{abstract}
The generation of optical vortex beams is pivotal for a myriad of
applications, encompassing optical tweezing, optical communications,
and quantum information, among others. The metasurface-based approach
has realized significant advancements in vortex production, utilizing
either dynamic or geometric phases. The dynamic design exhibits indifference
to the polarization state of incident light, while the geometric design
is inextricably tied to it. In the study, we put forth the proposition
that combining dynamic and geometric phases could unlock the potential
of metasurface design in generating optical vortices. A hybrid design
that harnesses the combined dynamic and geometric phases can attain
the same objective while offering tunable functional control over
the polarization of light. We establish a correlation between the
structural parameters of metasurface and the topological charge of
the resulting vortices. The experimental results fully demonstrate
the design's flexibility and its effective control over the polarization
constraints of incident light. Our research uncovers the capacity
for vortex generation through the manipulation of hybrid phases introduced
by metasurfaces, indicating significant potential for the design of
optical devices and the future advancement of innovative optical applications.
\end{abstract}
\maketitle

\section{{Introduction}}

\medskip{}

Optical vortex beams are typically paraxial beams characterized by
their cylindrical symmetric propagation. Notably, the vortex beam's
center is a dark core, where the intensity is nonexistent and remains
so throughout its propagation \cite{Dennis_2009,Shen_2019}. The
wavefront of vortex beams exhibits a spiral-shaped distribution, resulting
in its wavevector with an azimuthal component. Consequently, due to
the transverse spatial phase distribution, photons acquire orbital
angular momentum (OAM) \cite{Molina_Terriza_2007,Franke_Arnold_2008,Bliokh_2015}.
Owing to these characteristics, vortex beams offer distinct advantages
in various fields, including optical trapping, quantum entanglement,
nonlinear optics, optical processing, and high-resolution microscopic
imaging \cite{Yao_2011,Shen_2019}. In practical applications, vortex
beams can be produced directly using active vortex laser generators
\cite{Miao_2016,Zhang_2020}, or more commonly, by utilizing external
discrete optical elements to facilitate conversion \cite{Wang_2018}.
However, the optical components utilized in these techniques are bulky
and non-planar, presenting challenges for optical integration in various
applications.

A metasurface is an artificial material comprised of single or multiple
sub-wavelength nanostructural units, strategically arranged to perform
specific functions \cite{Yu_2011,Meinzer_2014,Kamali_2018}. By carefully
designing the geometry and composition of the nanostructural unit,
one can effectively control the polarization state, amplitude, and
phase of a light wave at a sub-wavelength scale \cite{Yu_2014,Overvig_2019,Bai_2022}.
The introduction of metasurfaces enables the manipulation of light
fields by incorporating helical wavefronts that are phase-varied with
respect to the azimuthal angle. This manipulation endows the beam
with an orbital angular momentum, thereby facilitating its transformation
into an optical vortex. Incorporating metasurfaces into the engineering
of optical vortex generation devices not only satisfies current requirements
for miniaturization, portability, and precise polarization control
but also substantially overcomes the constraints inherent in conventional
vortex-generating techniques.

Utilizing metasurfaces, the helical wavefront is generated by introducing
a spiral spatial phase shift through various nanostructural units
\cite{Zhang_pv_2018,Wang_2020,Guo_2022}. The optical light will
invariably experience a phase shift as it passes through these nanostructural
units, which can be precisely manipulated by design. Theoretically,
the phase shift of an optical beam can be decomposed into two distinct
components: the dynamic phase and the geometric phase \cite{Berry_1987,Guti_rrez_Vega_2011}.
The dynamic phase denotes the phase retardation induced by the optical
element as a consequence of the beam's propagation. This phase shift
is independent of the polarization or spin states of the light, being
primarily influenced by the frequency of the incident light and the
refractive index of the material \cite{Khorasaninejad_2016}. Given
that the structural thickness of nano units typically falls below
the operating wavelength, the modulation of the dynamic phase is predominantly
accomplished by altering the geometric dimensions or the spacing of
the nano units, which in turn modifies the effective refractive index.
The geometric phase originates from the interaction between spin angular
momentum and OAM, exhibiting a significant dependence on the polarization
state of the incoming light \cite{Bliokh_2015_2,Devlin_2017}. In
metasurface design, the geometric phase is frequently induced by rotating
anisotropic nano units tailored for circularly polarized light \cite{Marrucci_2006,Guo_2022}.
The two distinct types of phases have significantly influenced the
design of numerous metasurfaces for optical vortex generation, each
exhibiting unique functional advantages based on their underlying
principles. Despite their innovative aspects, these designs are confronted
with intrinsic limitations, notably in the realm of nanoscale structural
fabrication precision and the stringent specifications for polarization
control. The geometric phase and the dynamic phase exhibit distinct
correlations with the geometric configuration of the metasurface.
Adjusting the nano units' dimensions and orientation creates a hybrid
phase, offering potential to expand design limits and achieve significant
advancements \cite{Arbabi_2015,Balthasar_Mueller_2017,Zhang_2021}.
Considering that structural modifications influence both phases, precise
dimension control is key in developing hybrid phases \cite{Wang_2022}.
Therefore, elucidating the underlying correlations is essential for
the optimization of such designs.

In this paper, we initially detail the methodology for delineating
the dynamic and geometric phases introduced by metasurfaces. Thereafter,
we engage in a quantitative analysis of the respective contributions
of these phases to the generation of vortex beams, leveraging the
combination of these two phases. Building upon these analyses, we
engineer the metasurface structure to produce identical vortex beams
through varied designs, and ultimately validate our theoretical concepts
and design innovations through experimental verification. Through
the synergistic integration of geometric and dynamic phases, we are
able to design metasurfaces that exhibit enhanced flexibility and
augmented functionality in the context of optical vortex generation.
This advancement lays the groundwork for the development of innovative
metasurface designs that incorporate a variety of materials and novel
micro-nanostructures, thereby significantly benefiting the performance
and capabilities of a multitude of optical devices.

\section{Results and discussion}

\smallskip{}

\begin{figure}[ht]
\centering{}\centering \includegraphics[width=7.5cm]{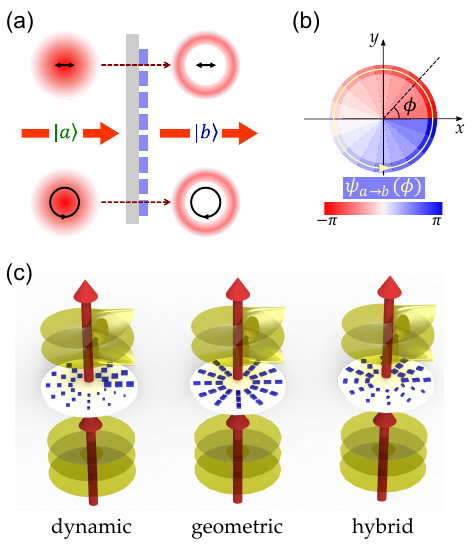} \caption{Optical vortex beams generated using metasurfaces. (a) A Gaussian
distributed planewave ($\left|a\right\rangle $) passes through metasurface
chip and is converted into a scalar vortex beam ($\left|b\right\rangle $)
with doughnut-like intensity distribution. (b) The spiral phase profile
is generated by the metasurface for the transformation from $\left|a\right\rangle $
to $\left|b\right\rangle $. (c) Left: Generating vortices with dynamic
phase gradient by varying the nano units size of metasurface. Center:
Generating vortices with geometric phase gradient by rotating the
nano units orientation for metasurface. Right: Generating vortices
with hybrid phase gradient by adjusting the size and orientation of
nano units. \label{fig1}}
\end{figure}

\subsection{Dynamic and geometric phases }

\begin{figure*}[t]
\centering{}\centering \includegraphics[width=15cm]{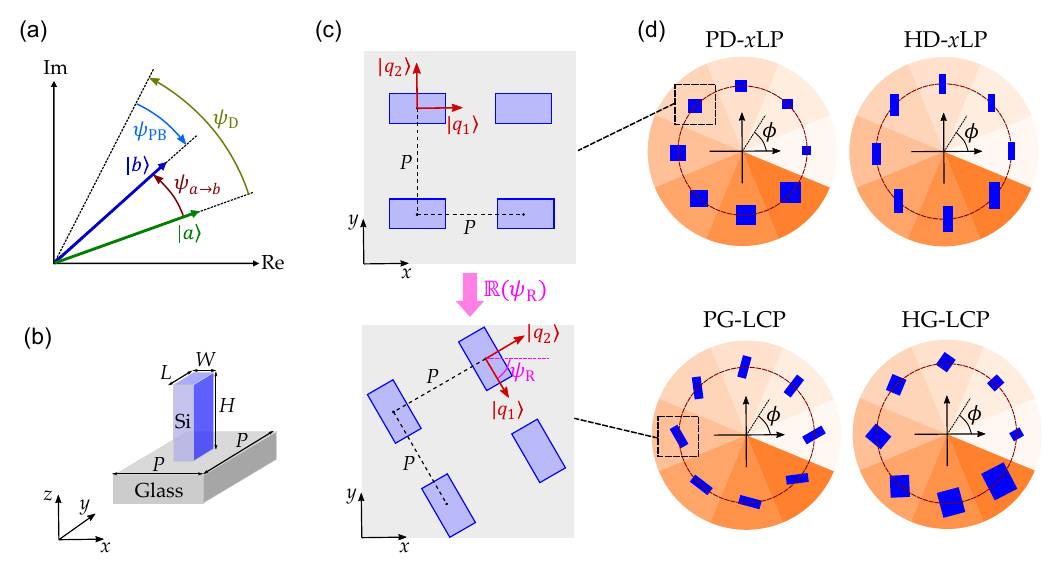} \caption{Phase shifts and configurations of metasurface. (a) Phase shifts from
$\left|a\right\rangle $ to $\left|b\right\rangle $ induced by metasurface.
(b) A nano-unit of the metasurface. (c) A uniform metasurface is achieved
utilizing a standard arrangement (upper) and an alternate, rotated
arrangement (lower). (d) Four metasurface designs, each based on an
eight-sector configuration, are engineered to generate scalar vortex
beams. The upper pair of metasurfaces is engineered to convert $x$-polarized
light, while the lower pair is designed to transform LCP light. \label{fig2}}
\end{figure*}

Firstly, we examine the methodologies for introducing two distinct
types of phase shifts. Let $\left|a\right\rangle $ represent the
input polarized light state; upon interaction with the metasurface,
this light is transformed into the output polarized state $\left|b\right\rangle $,
as illustrated in \figref{fig1}a. In accordance with the Pancharatnam
connection, the phase shift between this two states can be indicated
as $\psi_{a\rightarrow b}=\arg\left(\left\langle a|b\right\rangle \right)$
\cite{Guti_rrez_Vega_2011}. We can use Jones matrix $\mathbb{J}$
to represent optical elements, so the transformation process can be
described as $\left|b\right\rangle =\mathbb{J}\text{\ensuremath{\left|a\right\rangle }}$.
Assuming the $2\times2$ matrix $\mathbb{J}$ has two orthogonal eigenpolarization
states $\left|q_{1}\right\rangle $, $\left|q_{2}\right\rangle $
and the corresponding eigenvalues $\mu_{1}$, $\mu_{2}$. This implies
that, for input light with eigenpolarization $\left|q_{1,2}\right\rangle $,
the metasurface imparts a normalized transmittance of $\left|\mu_{1,2}\right|$
and a phase shift of $\arg(\mu_{1,2})$, constituting the eigen-responses.
Generally, the phase shift $\psi_{a\rightarrow b}$ between the input
$\left|a\right\rangle $ and the output $\left|b\right\rangle $ can
be divided into two components, 

\begin{equation}
\psi_{a\rightarrow b}=\psi_{\mathrm{D}}+\psi_{\mathrm{PB}},\label{eq:PhaseAtoB}
\end{equation}
where $\psi_{\mathrm{D}}$ denotes the dynamic phase and $\psi_{\mathrm{PB}}$
is named as the geometric phase, also known as the Pancharatnam-Berry
(PB) phase \cite{Berry_1987,Guti_rrez_Vega_2011,Mart_nez_Fuentes_2012}.
For a lossless conversion, there is $\psi_{\mathrm{D}}=\arg(\mu_{1}\mu_{2})/2$,
indicating the expected dynamic phase that the beam acquires upon
transmission through the metasurface. Further, $\psi_{\mathrm{PB}}$
can be given by

\begin{equation}
\psi_{\mathrm{PB}}=\arg\left[\cos\psi_{-}+\mathrm{i}\sin\psi_{-}\exp\left(\mathrm{i}\psi_{\mathrm{qa}}\right)\right],\label{eq:PBphase}
\end{equation}
where $\psi_{-}=\arg(\mu_{1}\mu_{2}^{*})/2$ and $\psi_{\mathrm{qa}}=\arg\left(\mathbf{Q}\cdot\mathbf{A}\right)$
with $\mathbf{Q}$ and $\mathbf{A}$ are the Stokes vectors corresponding
to $\left|q_{1}\right\rangle $ and $\left|a\right\rangle $, and
the detailed deduction can be found in Supporting Information. Owing
to the inclusion of the $\mathbf{Q}\cdot\mathbf{A}$ term, $\psi_{\mathrm{PB}}$
is associated with both the eigenstate of the metasurface and the
polarization state of the input light. Figure \fref{fig2}a illustrates
this relationship by depicting the phase shifts on the complex plane.

For the generation of an OAM light beam utilizing metasurfaces, a
spiral phase shift along the azimuthal angle $\phi$ is necessitated.
As shown in \figref{fig1}b, the phase shift $\psi_{a\rightarrow b}\left(\phi\right)$
facilitates the generation of a vortex beam, wherein the OAM charge
is associated with the gradient $\partial\psi_{a\rightarrow b}/\partial\phi$.
According to equation (1), the total OAM charge, denoted as $C_{\mathrm{Tot}}$,
can be bifurcated into two components: the dynamical contribution
($C_{\mathrm{D}}\propto\partial\psi_{\mathrm{D}}/\partial\phi$) and
the geometrical contribution ($C_{\mathrm{PB}}\propto\partial\psi_{\mathrm{PB}}/\partial\phi$),
thereby establishing the relationship $C_{\mathrm{Tot}}=C_{\mathrm{D}}+C_{\mathrm{PB}}$
\cite{Zhang_2018,Zhang_2015}. Using equation (1), by maintaining
$\psi_{\mathrm{PB}}$ as a constant value, we can achieve a pure dynamic
phase gradient by adjusting $\arg(\mu_{1}\mu_{2})$. This approach
is termed the pure-dynamic design methodology. Alternatively, by sustaining
$\psi_{\mathrm{D}}$ as a constant value, we can accomplish a pure
geometric phase gradient by varying $\psi_{\mathrm{PB}}$. This variation
can be achieved by modulating $\arg(\mu_{1}\mu_{2}^{*})$ and $\arg\left(\mathbf{Q}\cdot\mathbf{A}\right)$
as prescribed by equation (2). This methodology for vortex beam generation
epitomizes the pure-geometric design approach. In practice, the total
phase gradient can be actualized by concurrently modulating $\psi_{\mathrm{D}}$
and $\psi_{\mathrm{PB}}$ along $\phi$, integrating the dynamical
and geometrical contributions. This approach, which amalgamates both
phase components, can be denoted as the hybrid design methodology.

\smallskip{}

\subsection{General description for lossless metasurfaces }

As a fully polarized light beam traverses metasurface structures,
the specific nano-unit and its corresponding array induce a definitive
phase shift. The configuration of distinctly structured nano units
within an array can induce varying phase shifts. Consequently, a proper
design of the nano-unit combined with the strategic arrangement of
the array can generate a spiral phase distribution across the transverse
plane. For a lossless-transmission metasurface featuring nanofin arrays
in the $x-y$ plane, as shown in \figref{fig2}b and the upper panel
of \figref{fig2}c, there generally exist $x$($y$)-polarized eigenstates
denoted by $[1,0]^{\mathrm{T}}$ and $[0,1]^{\mathrm{T}}$. The corresponding
eigenvalues as $\mu_{1,2}=\exp(\mathrm{i}\varphi_{x,y})$, where $\varphi_{x,y}$
is the introduced phase shift for $x$($y$)-polarized light by the
nano-unit. Moreover, when a rotation angle $\psi_{\mathrm{R}}$ is
imposed on the array (see \figref{fig2}c), the eigenstates associated
with the nano-unit will be transformed to $\left|q_{1}\right\rangle =\mathbb{R}(\psi_{\mathrm{R}})[1,0]^{\mathrm{T}}$
and $\left|q_{2}\right\rangle =\mathbb{R}(\psi_{\mathrm{R}})[0,1]^{\mathrm{T}}$,
where $\mathbb{R}$ is the two-dimensional rotation matrix. For this
rotated nano-unit, the Jones matrix $\mathbb{J}$ can be expressed
as

\begin{widetext}

\begin{equation}
\mathbb{J}=\mathrm{e}^{\mathrm{i}\psi_{\mathrm{D}}}\left[\begin{matrix}\cos\left(\frac{\psi_{\mathrm{B}}}{2}\right)+\mathrm{i}\sin\left(\frac{\psi_{\mathrm{B}}}{2}\right)\cos(2\psi_{\mathrm{R}}) & \mathrm{i}\sin\left(\frac{\psi_{\mathrm{B}}}{2}\right)\sin(2\psi_{\mathrm{R}})\\
\mathrm{i}\sin\left(\frac{\psi_{\mathrm{B}}}{2}\right)\sin(2\psi_{\mathrm{R}}) & \cos\left(\frac{\psi_{\mathrm{B}}}{2}\right)-\mathrm{i}\sin\left(\frac{\psi_{\mathrm{B}}}{2}\right)\cos(2\psi_{\mathrm{R}})
\end{matrix}\right]\label{eq:JonesMatrix}
\end{equation}

\end{widetext}

\noindent where $\psi_{\mathrm{D}}=(\varphi_{x}+\varphi_{y})/2$ is
the introduced dynamic phase and $\psi_{\mathrm{B}}=\varphi_{x}-\varphi_{y}$
denotes the birefringent phase difference between transmissions of
the two eigenstates \cite{Balthasar_Mueller_2017,Zhang_2018}. Building
upon the preceding discussion, there are three methodologies to accomplish
the spiral phase profile for the generation of scalar vortex beams.
In the generations, the dynamic contribution to the OAM charge is
associated with $\partial\psi_{\mathrm{D}}/\partial\phi$, whereas
the geometric contribution, represented by $\partial\psi_{\mathrm{PB}}/\partial\phi$,
is linked to both $\psi_{\mathrm{B}}(\phi)$ and $\psi_{\mathrm{R}}(\phi)$,
since $\psi_{-}=\psi_{\mathrm{B}}/2$ and $\mathbf{Q}\cdot\mathbf{A}$
depends on the rotated eigenstates $\left|q_{1}\right\rangle $. 

In the pure-dynamic design approach utilizing metasurfaces, it is
imperative to maintain the birefringent phase difference at a constant
value. A prevalent strategy involves employing structures characterized
by circular symmetry, such as cylinders or circular holes \cite{Sun_2014},
to ensure that $\psi_{\mathrm{B}}=0$. For the unit founded on a nanofin
structure with $C_{2}$ symmetry, it is feasible to fine-tune $\mu_{1}$
and $\mu_{2}$ by modulating the dimensions of $W$ and $L$ (see
\figref{fig2}b). In order to realize $\psi_{\mathrm{B}}=0$, squared
nanofins (with $C_{4}$ symmetry) are employed to induce an identical
phase shift in both eigen-responses. Furthermore, by maintaining $\psi_{\mathrm{R}}(\phi)$
as a constant, $\arg\left(\mathbf{Q}\cdot\mathbf{A}\right)$ becomes
fixed, resulting in $C_{\mathrm{PB}}=0$. Subsequently, imposing a
spiral profile on $\psi_{\mathrm{D}}(\phi)$ by adjusting $\varphi_{x}+\varphi_{y}$
of the nano-unit is required. The OAM charge generated is determined
solely by $C_{\mathrm{Tot}}=C_{\mathrm{D}}$. 

\begin{figure}[b]
\centering{}\centering \includegraphics[width=8.5cm]{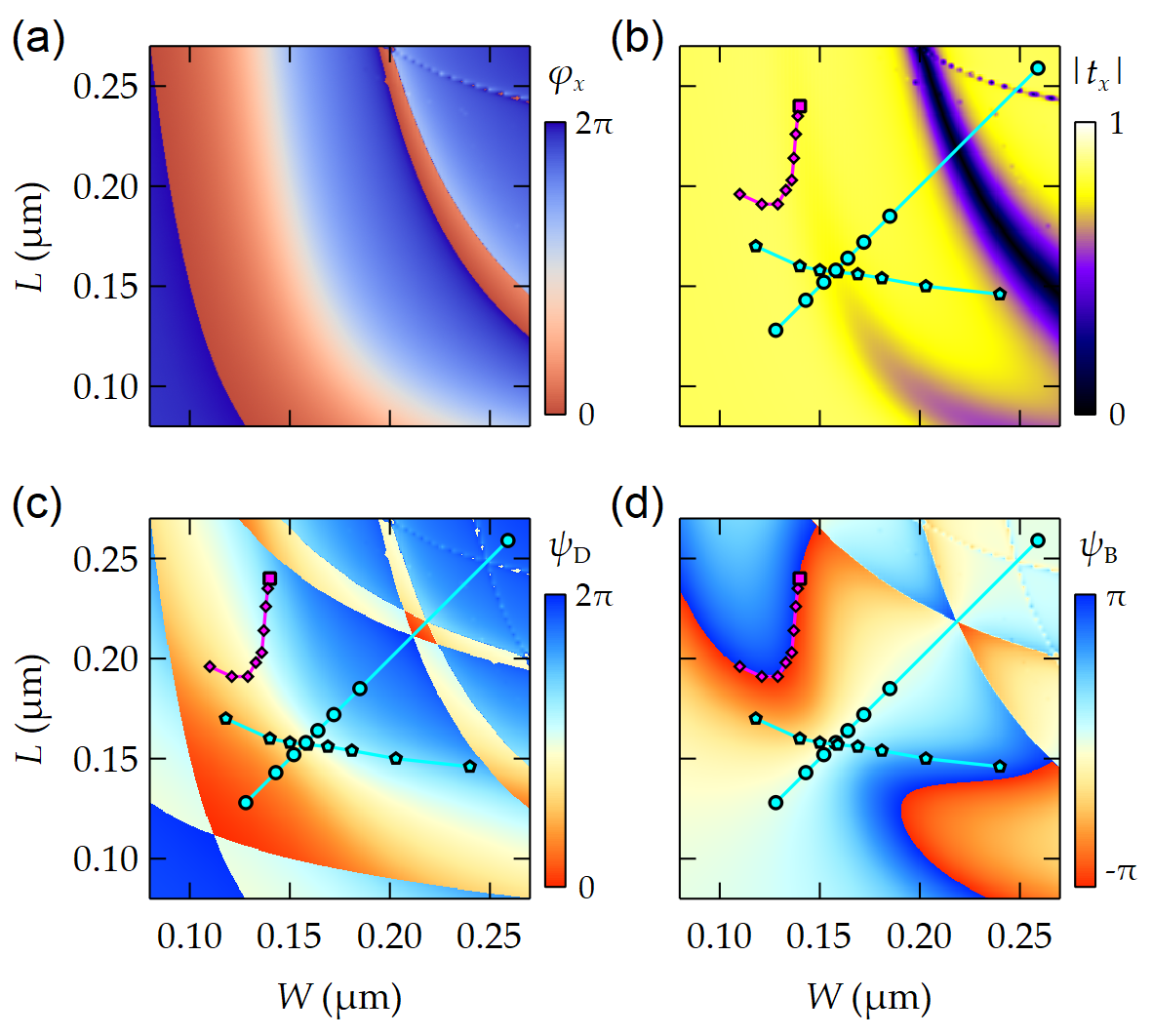} \caption{Basic optical features of uniform metasurfaces. (a) The phase shift
$\varphi_{x}$ and (b) transmittance $t_{x}$ exhibit dependencies
on the nanofin dimensions ($W$, $L$) for $x$-polarized light upon
transmission through a uniform metasurface. The $y$-polarized responses
can be obtained by exchanging the axes of $W$ and $L$ in (a) and
(b). Upon calculating the phase shifts for $x$- and $y$-polarized
lights, (c) the dynamic phase $\psi_{\mathrm{D}}$ and (d) the birefringent
phase difference $\psi_{\mathrm{B}}$ are determined and illustrated.
In (b-d), the eight circular markers correspond to parameters of PD-$x$LP,
while the eight pentagonal markers denote parameters of HD-$x$LP.
The square marker indicates parameters for PG-LCP, and the eight diamond
markers signify parameters for HG-LCP. \label{fig3}}
\end{figure}

For the pure-geometric design approach, by maintaining $\psi_{\mathrm{D}}$
and $\psi_{-}$ fixed as constants, we can manipulate the orientation
of the nano units to modify the value of $\arg\left(\mathbf{Q}\cdot\mathbf{A}\right)$,
thereby achieving pure PB-phase metasurfaces. This approach facilitates
the generation of phase gradients without altering the dimensions
of the nano units, accomplished solely by adjusting the orientation
angle \cite{Chen_2018,Huo_2020,Karimi_2014}. This technique is straightforward
and commonly employed in the design of geometric phase metasurfaces.
Specially, when $\psi_{-}$ equals to $\pi/2$ (i.e., $\psi_{\mathrm{B}}=\pi$)
and the input light is circularly polarized, the resultant output
is also a circularly polarized light with the opposite chirality.
In this transformation, we can derive a relationship where $\psi_{a\rightarrow b}=2\psi_{\mathrm{R}}$.
This straightforward link facilitates the extensive utilization of
circularly polarized light as an ideal input for PB-phase based metasurfaces.
For the nano-unit equipped with a nanofin, particular dimensions can
be chosen to set $\psi_{\mathrm{B}}=\pi$, followed by rotating the
nano-unit to induce a phase gradient along $\phi$. The resulting
OAM charge can be attributed $C_{\mathrm{Tot}}=C_{\mathrm{PB}}=2\partial\psi_{\mathrm{R}}/\partial\phi$. 

To demonstrate the efficacy of the hybrid design approach, our objective
is to achieve an equivalent outcome: generating an identical vortex
beam for a specified input light. In this study, we conducted two
hybrid design comparisons against the conventional pure-dynamic and
pure-geometric designs, respectively. For the pure-dynamic designed
metasurface, the total phase gradient $\partial\psi_{a\rightarrow b}/\partial\phi$
is solely derived from the dynamic phase profile $\psi_{\mathrm{D}}(\phi)$.
For comparative purposes, in the hybrid design, we diminish $\partial\psi_{\mathrm{D}}/\partial\phi$
while enhancing $\partial\psi_{\mathrm{PB}}/\partial\phi$ to maintain
a constant $\partial\psi_{a\rightarrow b}/\partial\phi$. This design
requirement can be readily fulfilled by meticulously adjusting $\varphi_{x}+\varphi_{y}$
and $\varphi_{x}-\varphi_{y}$ along $\phi$, while ensuring $\psi_{\mathrm{R}}(\phi)=0$.
Similarly, for the pure-geometric designed metasurface, the phase
gradient $\partial\psi_{a\rightarrow b}/\partial\phi$ is entirely
produced by the geometric phase profile $\psi_{\mathrm{PB}}(\phi)$.
For comparative analysis, we can diminish $\partial\psi_{\mathrm{PB}}/\partial\phi$
and augment $\partial\psi_{\mathrm{D}}/\partial\phi$ to maintain
a constant $\partial\psi_{a\rightarrow b}/\partial\phi$ in the hybrid
design. This manipulation can be accomplished by concurrently adjusting
$\varphi_{x}+\varphi_{y}$ and $\psi_{\mathrm{R}}$ along $\phi$,
with $\varphi_{x}-\varphi_{y}$ set to $\pi$. 

\smallskip{}

\subsection{Design and simulation}

\begin{figure*}[t]
\centering{}\centering \includegraphics[width=17cm]{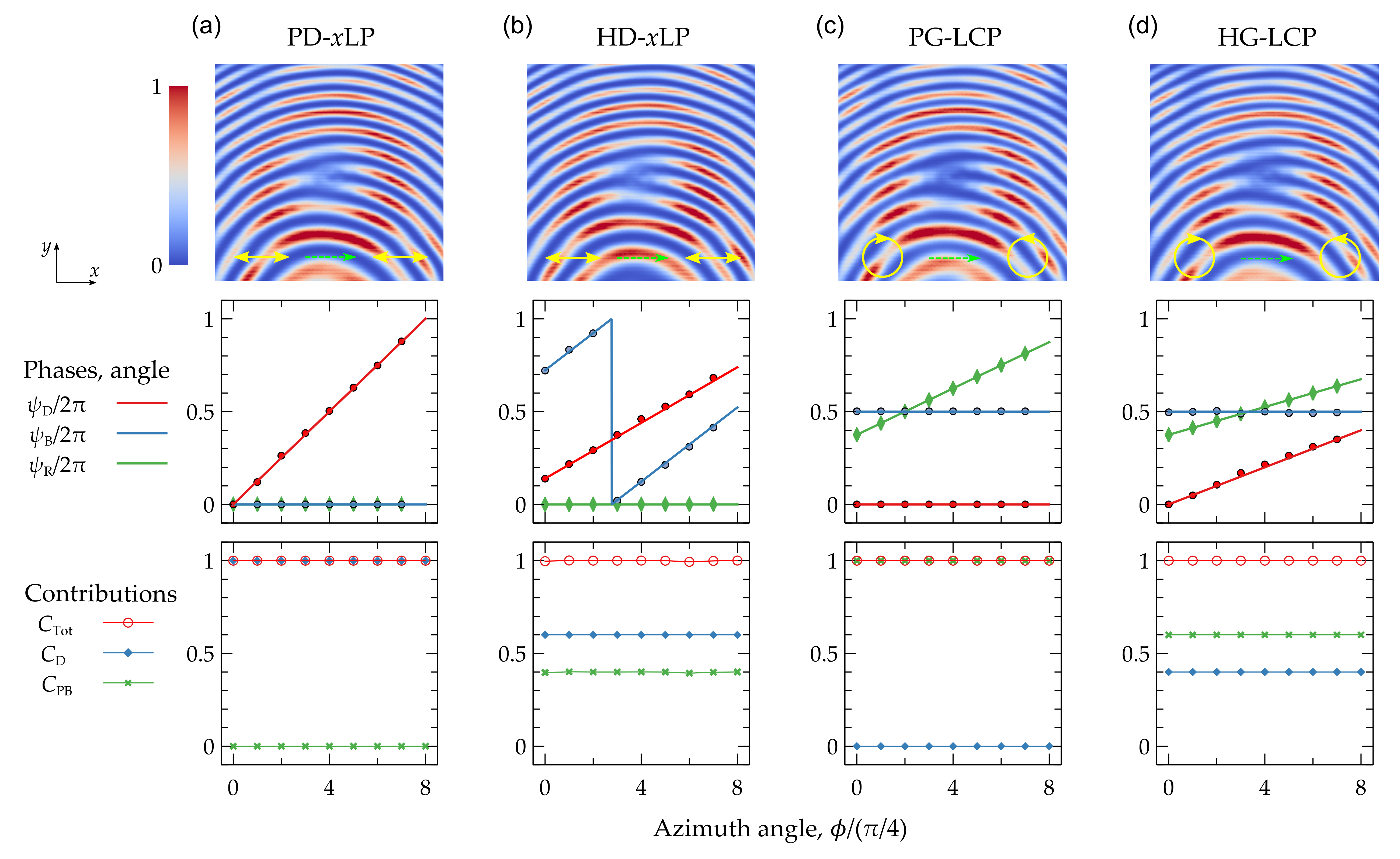} \caption{Four distinct metasurface configurations are engineered to produce
optical vortex beams. Metasurfaces utilizing (a) pure-dynamic phase
and (b) hybrid phase to convert an $x$-polarized planewave to an
$x$-polarized vortex beam. Metasurfaces employing (c) pure-geometric
phase and (d) hybrid phase to convert a LCP planewave to a RCP vortex
beam. For (a-d), the first row displays the calculated interference
patterns between a conventional Gaussian beam and the generated vortex
beam, where there is a shift between the centers of two beams, under
the ideal predefined incident polarization. There is a displacement
between the centers of the two beams in the interference. The second
row presents the designed phase set $\{\psi_{\mathrm{D}},\psi_{\mathrm{B}},\psi_{\mathrm{R}}\}$
of the nano-unit relative to the azimuth angle. Here, solid lines
represent the desired values, while the marked points indicate the
actual values of the selected nano-unit in each of the eight sectors.
The third row illustrates the contributions of the two types of phase
gradients to the OAM charge. \label{fig4}}
\end{figure*}

In our demonstration, the unit structure is made of a silicon nanofin
on a glass substrate, as shown in \figref{fig2}b. In order to determine
the eigen-responses of the nano unit array, we employ the Rigorous
Coupled-Wave Analysis (RCWA) to compute the transmitted eigen-responses
of periodic array structures upon the incidence of $x-$ and $y-$polarized
plane waves. In our design and simulation, the pitch of the nano units
was maintained at 400 nm in both directions. With a nanofin height
of 400 nm, by varying the width ($W$) and length ($L$) of nanofin,
the simulated $x$-polarized eigen-responses of $\varphi_{x}$ and
$t_{x}$ are displayed in \figsref{fig3}a  and \fref{fig3}b,
respectively. These two parameters respectively denote the phase shift
and transmittance experienced by $x$-polarized light upon passing
through a uniform array. Owning to the $C_{2}$ symmetry of the nanofin,
the $y$-polarized eigen-responses of $\varphi_{y}$ and $t_{y}$
can be easily determined by merely switching the coordinate axes in
\figsref{fig3}a  and \fref{fig3}b. Utilizing the simulated maps
of $\varphi_{x}$ and $\varphi_{y}$, the corresponding maps for $\psi_{\mathrm{D}}$
and $\psi_{\mathrm{B}}$ are derived and displayed in \figsref{fig3}c
 and \fref{fig3}d, respectively. To generate vortex beams, a spiral
spatial phase profile must be incorporated into the cross-section
that is orthogonal to the direction of propagation. In our experimental
demonstrations, the spiral phase profile is constructed by segmenting
the structure into eight sectors along the azimuthal angle $\phi$,
each characterized by discrete phase increments, as shown \figref{fig2}d.
Within each sector, a uniform array of nano units is meticulously
arranged to facilitate the desired phase shifts. To approximate a
lossless condition, all nano units are chosen for their high transmittance.

\begin{figure*}
\centering{}\centering \includegraphics[width=16cm]{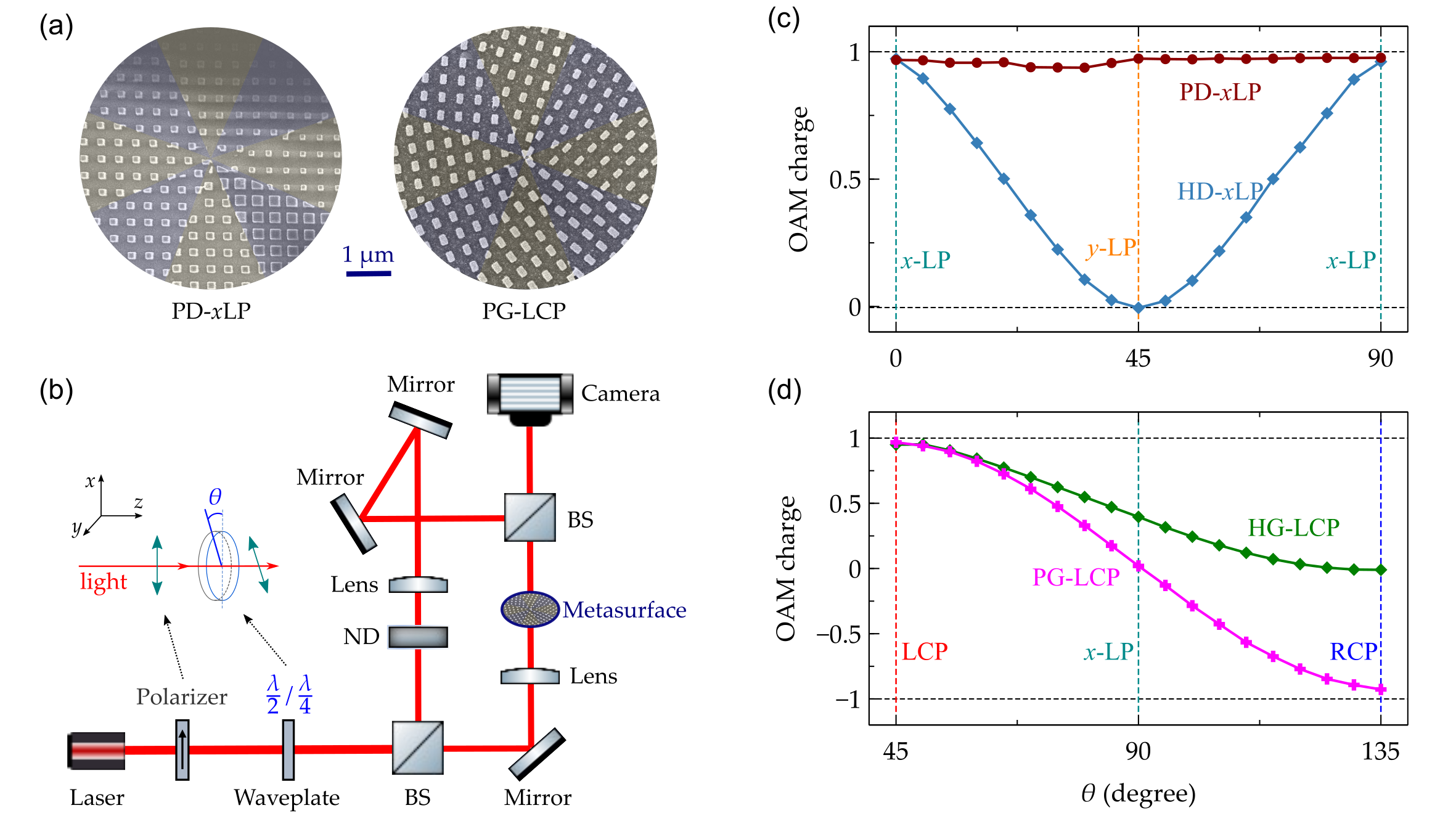} \caption{Fabricated metasurfaces and measurement setups. (a) Scanning electron
microscope (SEM) images of the fabricated samples of PD-$x$LP and
PG-LCP designs. (b) The proposed measurement apparatus for the optical
vortex beams produced by the designed metasurfaces relies on the interference
of a reference beam with the generated vortex beam to yield the interference
pattern. The beam splitter (BS) facilitates either the division or
combination of light beams, while the neutral density filter (ND)
is utilized to equilibrate the powers of the two beams, ensuring prominent
interference fringes. The inset illustrates a polarizer accompanied
by a $\lambda/2$ ($\lambda/4$) waveplate, employed to alter the
polarization state of the incident light. (c) The average OAM charge
of optical vortex beams generated by PD-$x$LP and HD-$x$LP designs
is calculated and graphically represented at different directions
of incident linear polarizations, which are adjusted using the polarizer
and the rotated $\lambda/2$ waveplate. (d) The average OAM charge
of vortex beams produced by PG-LCP and HG-LCP designs is computed
and plotted at distinct incident polarization states (from LCP to
RCP), which are modified using the polarizer and the rotated $\lambda/4$
waveplate. \label{fig5}}
\end{figure*}

As described previously, we will undertake four metasurface designs
to produce scalar vortex beams with an OAM charge of 1. The first
design, denoted as PD-$x$LP, involves the generation of a pure-dynamic
vortex beam. In this case, the sampled $\psi_{\mathrm{D}}(\phi)$
values for the eight sectors are incrementally increased, while $\psi_{\mathrm{B}}(\phi)$
is set to $0$ for all nano units without any rotation, ensuring that
$\psi_{\mathrm{R}}(\phi)=0$. In comparison, the hybrid design referred
to as HD-$x$LP integrates the spiral phase by sampling both $\psi_{\mathrm{D}}(\phi)$
and $\psi_{\mathrm{B}}(\phi)$, applying incremental steps to the
nano units across the eight sectors, while $\psi_{\mathrm{R}}(\phi)$
is maintained at $0$. The schematics for both designs are presented
in the first row of \figref{fig2}d. Both of PD-$x$LP and HD-$x$LP
designs have the capability to convert an $x$-polarized plane wave
into a scalar vortex beam with an OAM charge of 1. The pure-geometric
design, denoted as PG-LCP, is illustrated for converting a LCP planewave
into a vortex beam with the opposite polarization chirality. The gradient
of geometric phase is accomplished by rotating the uniform nano-units,
which have $\psi_{\mathrm{B}}(\phi)=\pi$, to different angles $\psi_{\mathrm{R}}(\phi)$
within each sector. For comparison, the hybrid design, denoted as
HG-LCP, is implemented by incorporating varied $\psi_{\mathrm{D}}(\phi)$
through the adjustment of eight nano-unit structures, while maintaining
$\psi_{\mathrm{B}}(\phi)=\pi$. The illustrations of the two designs
are presented in the second row of \figref{fig2}d. Both the PG-LCP
and HG-LCP designs have the capability to convert an LCP plane wave
into an RCP vortex beam with an OAM charge of 1. 

Taking these considerations into account, the dimensions of the nanofins
for the four designs are selected and marked in \figsref{fig3}{b-d}.
The corresponding set of $\{\psi_{\mathrm{D}},\psi_{\mathrm{B}},\psi_{\mathrm{R}}\}$
for each nano-unit are summarized and plotted in the middle panels
of \figsref{fig4}{a-d} for the four designs, respectively. Furthermore,
the contributions for the OAM charge can be evaluated and displayed
in the bottom panels of \figsref{fig4}{a-d}. With the simulation
results of each nano-unit shown in \figsref{fig3}{a-d}, the transformed
far field can be calculated with the corresponding input light beams.
To confirm the OAM charge of the generated vortices, the simulated
interference patterns of the vortex beams with a conventional Gaussian
beam are displayed in the top panels of \figsref{fig4}{a-d}. The
superposition of a vortex beam and a plane wave results in an interference
pattern characterized by a fork-like bifurcation at the vortex core
\cite{Liu_2018}. The specific morphology of this bifurcation is
correlated with the OAM charge of the beam. As patterns shown in \figsref{fig4}{a-d},
the red-colored toroidal contour is a distinctive characteristic of
vortex beams, with a fork-shaped pattern at the center, featuring
two distinct branches that signify an OAM charge of 1. 

\begin{figure}[t]
\centering{}\centering \includegraphics[width=8.5cm]{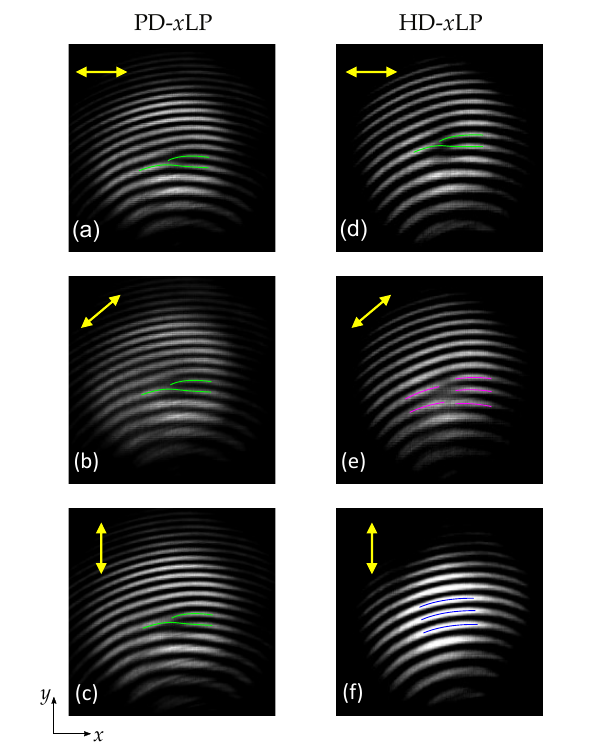} \caption{Measured interference patterns for optical vortex beams generated
by PD-$x$LP and HD-$x$LP metasurfaces at different directions of
incident linear polarizations: (a,d) $x$-polarized, (b,e) $45^{\circ}$linearly
polarized, and (c,f) $y$-polarized. \label{fig6}}
\end{figure}

\smallskip{}

\subsection{Experimental results }

The designed samples were fabricated using a borosilicate glass substrate
with a thickness of 1~mm, followed by the evaporation of a 400~nm
silicon film. Subsequently, silicon nanofins were patterned within
a 0.5~mm diameter circle through electron-beam lithography and a
dry etching process. Figure \fref{fig5}a shows scanning electron
microscope (SEM) images of two fabricated samples corresponding to
the PD-$x$LP and PG-LCP designs. A Mach-Zehnder interferometric system,
constructed for characterization purposes, is illustrated in \figref{fig5}b.
The light beam is produced by a laser with a wavelength of 780~nm
and is adjusted to $x$-polarized light using a polarizer. Subsequently,
a half-wave ($\lambda/2$) or quarter-wave ($\lambda/4$) waveplate
is employed to transform the $x$-polarized light into linearly, elliptically,
or circularly polarized light. The light beam is subsequently split
into two, with one portion passing through the metasurface chip. These
two beams are ultimately recombined, generating interference patterns
that are captured by a charge-coupled device (CCD) camera. As shown
in the inset of \figref{fig5}b, by changing the angle $\theta$
between the optical axes of polarizer and $\lambda/2$ (or $\lambda/4$)
waveplate, any desired linearly polarized (or circularly polarized)
light can be obtained. Utilizing these polarization modifications,
we are able to confirm the production of vortex beams induced by the
metasurfaces and further assess the reliance of various metasurfaces
on the polarization states of the incident light.

In the configuration involving the polarizer and the $\lambda/2$
waveplate, the orientation of linear polarization undergoes a rotation
of $2\theta$ corresponding to the rotation $\theta$ of the $\lambda/2$
waveplate. For the pure-dynamic metasurface (PD-$x$LP), the rotation
of the $\lambda/2$ waveplate exerts a minimal impact on the outcomes,
which is consistent with its polarization-independent properties,
as shown in \figref{fig5}c. The light field and OAM charge of the
generated vortex beams remain unaffected by the polarization state
of the input light, with their interference fringes displayed in \figsref{fig6}{a-c}.
In the case of the hybrid design (HD-$x$LP), a distinct polarization
dependence is observed, as illustrated by the calculated OAM charge
shown in \figref{fig5}c. An optimal vortex beam is generated when
the incident light is $x$-polarized. When the incident light is $y$-polarized,
the output remains a planewave, and the interference fringes manifest
as alternating bright and dark stripes. These findings are confirmed
by experimental results presented in \figsref{fig6}{d-f} (also
see Supporting Information). It is crucial to highlight that, for
both the PD-$x$LP and HD-$x$LP designs, identical vortex beams can
be produced using $x$-polarized incident light.

\begin{figure}[t]
\centering{}\centering \includegraphics[width=8.5cm]{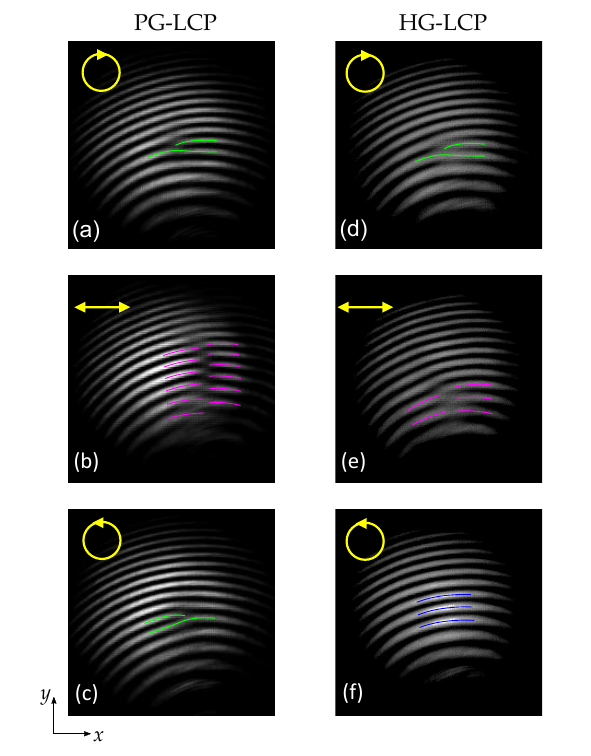} \caption{Measured interference patterns for optical vortex beams generated
by PG-LCP and HG-LCP metasurfaces at different incident polarization
states: (a,d) LCP, (b,e) $x$-polarization, and (c,f) RCP. \label{fig7}}
\end{figure}

Typically, the geometric phase exhibits high sensitivity to the polarization
state of the incident light. As the light traverses the polarizer
and the $\lambda/4$ waveplate, its polarization can be transformed
into linear, elliptical, or circular polarization, depending on the
rotation $\theta$ of the $\lambda/4$ waveplate. According to the
calculations illustrated in \figref{fig5}d, the OAM charge of the
pure-geometric design (PG-LCP) is heavily reliant on the polarization
state of light. Specifically, the sign of the OAM charge is inverted
for input light of opposite chirality, reflecting the conjugate symmetry
inherent in the relationship between the input and output fields.
From the interference fringes shown in \figsref{fig7}a and \fref{fig7}c,
it is observed that input light of distinct chiralities results in
the formation of interference fringes oriented in opposite directions.
However, upon linear polarization of the input light, the output light
undergoes transformation into a vector vortex beam (see Supporting
Information). Upon interference with circularly polarized reference
light, an interlaced pattern of interference fringes is theoretically
produced and the experimental result is shown in \figref{fig7}b.
In the case of the hybrid design (HG-LCP), the OAM charge of the output
light remains associated with the incident polarization state (see
\figref{fig5}d). Nevertheless, compared to the pure geometric design,
the fluctuation in the OAM charge is attenuated, and no distinct reversal
is observed. Experimental findings suggest that a well-defined vortex
beam is exclusively produced when the input light possesses an LCP
state (see \figref{fig7}d). Conversely, employing RCP light results
in an output light with a negligible topological charge, culminating
in an interference pattern characterized by alternating bright and
dark stripes (see \figref{fig7}f). Consequently, both the PG-LCP
and HG-LCP metasurfaces are capable of executing identical transformations
for LCP input light. However, the extent of reliance on the incident
polarization state varies, as the dynamic contribution within the
hybrid design can effectively mitigate polarization dependence.

In this study, we showcased that metasurfaces with various designs
can produce scalar vortex beams. For a particular transformation,
aside from the conventional methods employing pure-dynamic or pure-geometric
generation of the spiral phase profile, there exists substantial potential
in designs that integrate both dynamic and geometric contributions.
These hybrid designs can overcome certain constraints regarding the
choice of materials and forms for nano-units. Furthermore, transformations
based on hybrid designs offer distinct advantages: In the case of
a pure-dynamic design, there is no selectivity towards the polarization
state of the input light. However, the hybrid design can introduce
a polarization filter, and its sensitivity or bandwidth can be customized
by adjusting the geometric contribution. Conversely, the pure-geometric
design mandates a stringent requirement for the polarization state
of the input light; however, this constraint can be alleviated in
the hybrid design by incorporating the dynamic contribution. In the
experiment, we successfully generated a vortex beam with an OAM charge
of 1. However, there are no constraints on producing higher-order
optical vortices using hybrid designs; this can be accomplished by
expanding the number of sampling sectors and enhancing the requisite
phase gradient. Besides generating scalar vortex beams, hybrid designs
offer the advantage of producing vector vortices, thanks to the incorporated
geometric contributions.

\section{Conclusion }

\medskip{}
Metasurfaces can yield dynamic and geometric phases, both of which
are extensively employed to produce optical vortex beams. However,
in reality, by leveraging the combined dynamic and geometric phases,
innovative and diverse functionalities can be realized in the engineering
of optical vortex beam generation. In this study, the spiral phase
distribution is achieved through the customization of hybrid phases
to produce scalar vortex beams. Additionally, the OAM contributions
from the two phases can be precisely apportioned. Utilizing metasurfaces
that incorporate hybrid phases, we introduce an optimization strategy
for the engineering of nano units, addressing both structural and
functional dimensions. The hybrid design concept we introduce is not
restricted by the material, structure, or propagation manner of metasurfaces.
This principle opens up the possibility of employing hybrid phases
to engineer a variety of phase profiles that can facilitate new functionalities
and pave the way for innovative optical devices in the future.

\bigskip{}

\noindent \textbf{Funding}

\noindent This work is supported by the Fundamental Research Funds
for the Central Universities under Grant KG21008401.

\medskip{}

\noindent \textbf{Disclosures. }The authors declare no competing financial
interests.

\noindent \textbf{Data availability. }Data underlying the results
presented in this paper are not publicly available at this time but
may be obtained from the authors upon reasonable request.

\medskip{}

\noindent \textbf{Supporting Information. }Detailed theoretical derivation
of phase components, simulated transformations of the polarization
state from input to output, and experimental far-field patterns of
vortex beams and interference fringes at varying input polarization
states.

\bibliographystyle{naturemag}

\end{document}